\documentclass[onecolumn,notitlepage,altaffilletter,aip,rsi,numerical]{revtex4-1}
\usepackage{amsmath}
\usepackage{graphicx}				
\usepackage{url}					
\usepackage{amssymb}
\usepackage{natbib}
\begin{document}

\title{Oscillator metrology with software defined radio}
\author{Jeff A.\ Sherman}
\email{jeff.sherman@nist.gov}
\author{Robert J\"{o}rdens}
\email{rj@m-labs.hk}
\affiliation{National Institute of Standards and Technology, Division of Time and Frequency, Boulder, Colorado, USA}
\date{\today}

\begin{abstract}
Analog electrical elements such as mixers, filters, transfer oscillators, isolating buffers, dividers, and even transmission lines contribute technical noise and unwanted environmental coupling in time and frequency measurements. Software defined radio (SDR) techniques replace many of these analog components with digital signal processing (DSP) on rapidly sampled signals. We demonstrate that, generically, commercially available multi-channel SDRs are capable of time and frequency metrology, outperforming purpose-built devices by as much as an order-of-magnitude. For example, for signals at 10~MHz and 6~GHz, we observe SDR time deviation noise floors of about 20~fs and 1~fs, respectively, in under 10~ms of averaging. Examining the other complex signal component, we find a relative amplitude measurement instability of $3 \times 10^{-7}$ at 5~MHz. We discuss the scalability of a SDR-based system for simultaneous measurement of many clocks. SDR's frequency agility allows for comparison of oscillators at widely different frequencies. We demonstrate a novel and extreme example with optical clock frequencies differing by many terahertz: using a femtosecond-laser frequency comb and SDR, we show femtosecond-level time comparisons of ultra-stable lasers with zero measurement dead-time.
\end{abstract}
\maketitle

\section{Overview}
Time is best measured by counting periods of natural or manmade oscillators~\cite{jespersen1999sundials}. To maximize temporal resolution we must interpolate between integer periods, a task equivalent to determining an oscillator's phase. Consider two oscillators with frequency $f$, the periods of which can be counted as clocks. Their phase offset $\Delta\phi(t_k)$ (in radians) at a measurement epoch $t_k$ can be interpreted as a time offset~\cite{allan1975picosecond},
\begin{equation}
\label{eq:timephase}
\Delta T(t_k) = \frac{\Delta \phi(t_k)}{2 \pi f}.
\end{equation}
Resolving whether $\Delta T$ is stationary is the most sensitive method for detecting small frequency offsets or fluctuations between the oscillators and thus calibrating or characterizing them as clocks~\cite{walls1986measurements,levine2012invited}. For continuously running clocks, a linear drift in $\Delta T$ defines a (fractional) frequency offset between the oscillators $y = \left[\Delta T(t_2) - \Delta T(t_1)\right] /(t_2 - t_1)$ consistent with the notion that frequency is the rate of change of phase.

In this work, we briefly review existing high-resolution techniques for measuring $\Delta T$ of radio frequency oscillators. We introduce the software defined radio (SDR) concept in the context of time and frequency metrology, and describe basic demonstration experiments valid for many SDR implementations. Finally, we explore SDR's ability to compare oscillators at dissimilar frequencies and to scale to many-oscillator comparisons. One new SDR application is discussed in some detail: phase-coherent measurement of optical clocks via a femtosecond laser frequency comb.

\subsection{Radio techniques in oscillator metrology}
\begin{figure}
\begin{center}
\includegraphics[scale=0.9]{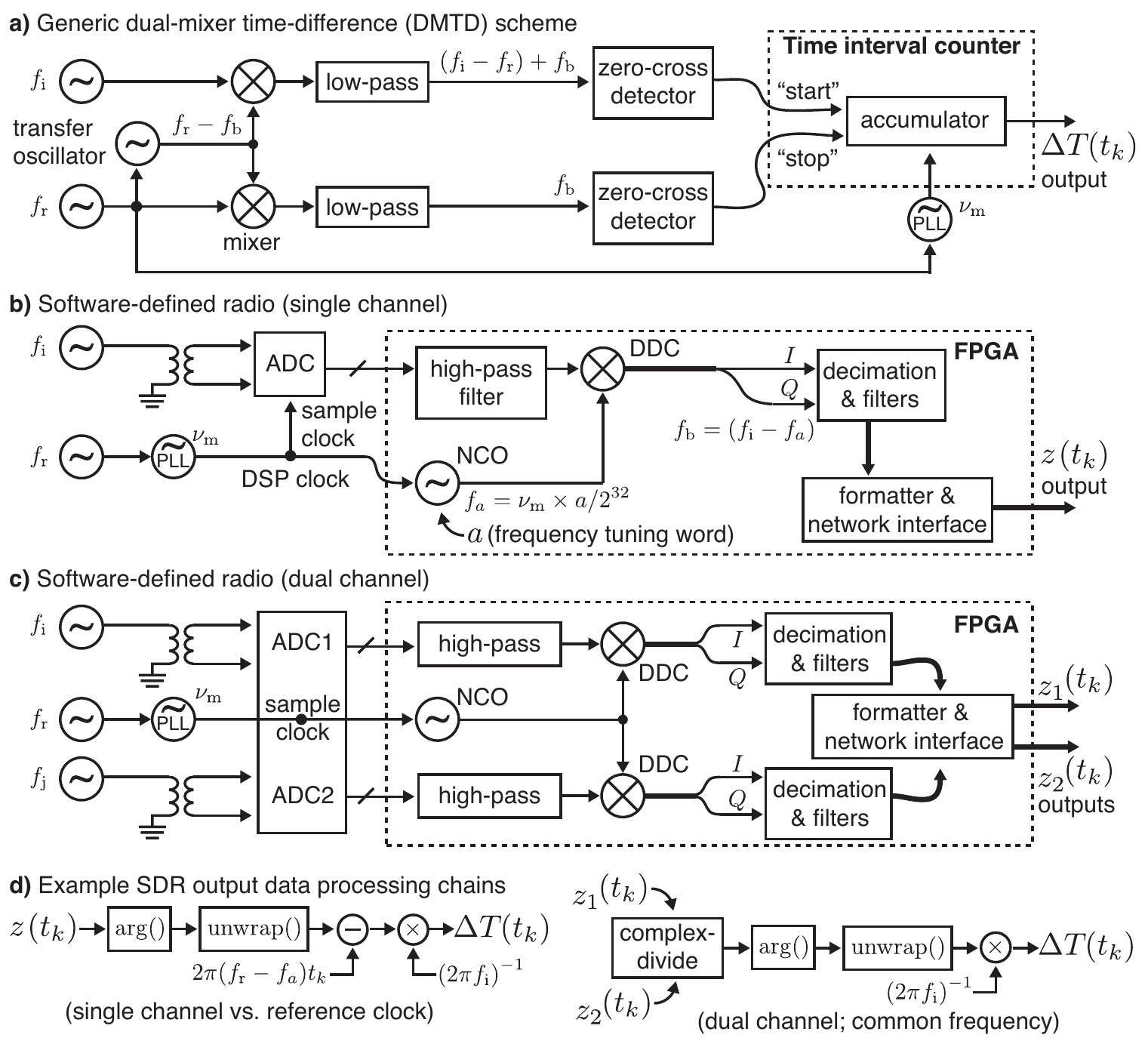}

\end{center}
\caption{Schematic comparison of a) generic dual-mixer time-difference (DMTD), and b) a software defined radio (SDR) described here. Dashed-lines surround digital processing sections. In both concepts, $f_\text{i}$ is the frequency of an oscillator under test. A reference oscillator $f_\text{r}$ disciplines a digital clock at fast frequency $\nu_\text{m}$ through a phase-locked-loop (PLL). Both methods gain resolution by spectrally shifting $f_\text{i}$ to a low frequency $f_\text{b}$; in SDR, the mixer analogue is digital downconversion (DDC) with a synthesized numerically-controlled oscillator (NCO). c) Two oscillators $f_\text{i}$ and $f_\text{j}$ are compared in two channels of a single ADC, suppressing noise due to the $\nu_\text{m}$ PLL, its reference tone $f_\text{r}$, and the ADC's aperture jitter. d) While DMTD directly outputs time-offset data; further processing is performed on the SDR sampled waveform $z(t_k)$ with a computer to determine time offset. We illustrate two simplified processing chains for single- and dual-channel measurements; see text for details.}
\label{fig:heterodyne_comparison}
\end{figure}

Though clock frequencies may be high, measurement bandwidth need not be for comparing oscillators $i$ and $r$ that are similar in frequency, $f_\text{i} \approx f_\text{r}$. Since clock oscillators are typically very stable, a signal at $f_\text{i} - f_\text{r}$ is both low in bandwidth and low in absolute frequency and therefore amenable to high-precision measurement. Such frequency translation is rooted in radio techniques---transmitters shift a signal of low- to moderate-bandwidth upwards to many megahertz or gigahertz for ease of wide-area propagation while receivers spectrally convert the signal back to its original band with no practical loss in information.

The widely applied dual-mixer time-difference (DMTD) technique~\cite{allan1975picosecond}, illustrated in Figure~\ref{fig:heterodyne_comparison}a, is an example of radio frequency translation applied to oscillator metrology. A \emph{transfer} oscillator is synthesized at $f_\text{r} - f_\text{b}$, slightly offset from \emph{input} and \emph{reference} oscillators $f_\text{i}$ and $f_\text{r}$:  ($|f_\text{i} - f_\text{r}| \ll f_\text{b} \ll f_\text{i,r}$). The transfer oscillator is mixed (multiplied) with both $f_\text{r}$ and $f_\text{i}$ tones, creating two signals with frequencies near $f_\text{b}$ after low-pass filtering. A time-interval counter (TIC) counts periods of a fast \emph{timebase} oscillator $\nu_\text{m}$, also disciplined by $f_\text{r}$, gated by high slew-rate zero-crossing detectors observing the two heterodyne products near $f_\text{b}$.

As a consequence of the spectral conversion, DMTD methods resolve $\Delta T \le 1$~ps accurately, despite no component possessing a bandwidth approaching $(\text{1 ps})^{-1} = 1$~THz. DMTD realizations often employ an offset frequency $1~\text{Hz} \le f_\text{b} \le 10~\text{Hz}$ and heterodyne factors $10^5 \le f_\text{r} / f_\text{b} \le 10^7$, so a TIC must only accurately resolve $\Delta T \sim 1~\mu$s between the relatively slow oscillations near $f_\text{b}$ to discern $(f_\text{r} / f_\text{b})^{-1}(1~\mu\text{s}) \sim1$~ps of oscillator time difference.

While DMTD techniques are already highly refined~\cite{sojdr2004optimization}, $\Delta T$ of 1~ps can be mimicked or masked by fluctuations of $\approx 200~\mu$m in electrical length or group delay, so even cables contribute to instability (the temperature dependence is of order 0.5 ps m$^{-1}$ K$^{-1}$). Analog components (including `digital' mixers) can contribute flicker-phase noise~\cite{phillips2000noise}, amplitude-to-phase-modulation conversion~\cite{cibiel2002noise}, sensitivity to interference (e.g., channel crosstalk, ground loops, wideband ambient rf), and coupling to the environment (e.g., temperature, humidity).  The TIC start- and stop-inputs require high-bandwidth, high slew-rate triggers, but the signals following the mixers are slow sinusoids. Zero-crossing detectors must therefore boost signal slew rates by $\sim 10^6$ while accurately preserving phase~\cite{brida2002high}. Without additional synthesis steps, DMTD requires $f_\text{r}$ and  $f_\text{i}$ to be very similar, and among a small set of frequencies compatible with the analog processing components. Mixer and filter non-linearity and frequency-dependent group delay complicate maintaining a whole-system $\Delta T$ calibration over arbitrary signal frequencies. Finally, DMTD schemes cannot resolve fluctuations over time scales shorter than $1/f_\text{b}$.

\subsection{Related work}
Some limitations in DMTD can be addressed by replacing certain analog processing steps with digital implementations. The TIC can be dramatically redesigned~\cite{prochazka2014note} with much higher effective $\nu_\text{m}$. One group~\cite{uchino2004frequency} replaced the TIC by digitizing the mixed and filtered signals at $f_\text{b}$ and later eliminated the mixers with high-speed direct sampling of the input signals~\cite{mochizuki2007frequency}. Others have replaced mixer-based spectral down-conversion with aliasing through under-sampling~\cite{romisch2011digital}. Early consideration of a direct-sampling system~\cite{landis2001new} very similar to the present work showed plausible limits due to quantization effects alone can be $\Delta T < 1~\text{fs } (\tau/1~\text{s})^{-1/2}$ for averaging intervals $\tau$. While high-speed samples can be processed entirely in software~\cite{ruppalt2010simultaneous} or with custom hardware~\cite{grove2004direct}, this work explores oscillator metrology using an inexpensive, commercially available, unmodified software defined radio (SDR). We note a similar approach for characterizing ADCs~\cite{cardenas2015simple}. We employ an ADC noise cancelation technique in the time domain, which perhaps is analogous to cross-spectral analysis~\cite{rubiola2010cross} in the frequency domain.

\section{Software defined radio}
\subsection{``Sample first, ask questions later''}
In SDR~\cite{mitola1995software}, signals of interest are sampled by a fast, high-resolution analog-to-digital converter (ADC) with little or no analog processing, amplification, or filtering. A numerically-controlled oscillator (NCO), computed synchronously with ADC sampling, takes the place of the local oscillator tone in analog radio reception. A digital multiplication of the sampled signal and NCO phasor performs the role of signal mixer. Filtering and sample rate decimation are also performed digitally, reducing noise bandwidth while conserving signal information. Here we focus on SDR receiver functions, but many SDRs are capable of transmission as well. Since the signal processing chain in SDR is highly-configurable, it has applications in radar, spread-spectrum and multiple-input multiple-output (MIMO) communication, and advanced protocol demodulation and simulation.

SDR seems to suffer a significant disadvantage: noise figures of high-speed ADCs are much worse than a collection of radio frequency filters, amplifiers, and mixers. On the other hand---especially in the context of precision metrology---analog components are subject to strict impedance matching requirements and exhibit long-term sensitivity to shock, vibration, supply voltage, temperature, humidity, aging, interference, and signal crosstalk. A low ADC signal-to-noise ratio (SNR) is at least amenable to averaging and process gain, while environmental sensitivities are more pernicious sources of stochastic noise and drift over long durations. In contrast, digital processing steps are stable, deterministic, and environmentally insensitive.

\subsection{Technical details}
\label{sec:technical_details}
At the time of writing, the techniques presented here ought to apply to SDRs from at least ten manufacturers. While we attempt to consider SDR generically, Figure~\ref{fig:heterodyne_comparison}b illustrates relevant components in the SDR receiver studied here (an Ettus USRP N210 except where noted~\cite{nist_disclaimer}). Field programmable gate array (FPGA) hardware description code and circuit schematics are available for inspection and customization~\cite{ettus_site,ettus_github}. A receiver daughterboard couples a ground-referenced input signal (1 to 250~MHz) into a differential ADC via a transformer. The ADC (Texas Instruments ADS62P44) has an analog input bandwidth of 450~MHz ($-3$dB), a full-scale range of $\pm 1$~V, and 14-bit resolution. The ADC specifications include an aperture jitter $t_\text{ap} = 150$~fs, a significant technical timing uncertainty between an idealized sample trigger and actuation of the converter's sample-and-hold circuitry. The sample timebase, a voltage-controlled crystal oscillator (VCXO) at $\nu_\text{m} =100$~MHz, drives the ADC sampling trigger and the FPGA's digital signal processing (DSP) pipeline. The SDR includes phase-locked loop  (PLL, bandwidth $\approx 3$~kHz) components, which we often use to discipline $\nu_\text{m}$ to a $+14$~dBm signal at $f_\text{r} = 10$~MHz derived from an active hydrogen maser. In our configuration, this SDR consumes about 10~W of dc power.

SDR's three important DSP tasks are frequency translation, filtering, and data decimation. After a high-pass filter suppresses the ADC's zero-offset, the input signal undergoes digital down-conversion (DDC), or frequency translation by an NCO tuned to 
\begin{equation}
f_a = \nu_\text{m} \times \frac{a}{2^{32}},
\label{eq:ftw}
\end{equation} 
where $a$ is an integer $0 \le a < 2^{32}$. As in direct digital synthesis (DDS), a phase register accumulates the frequency-tuning word $a$ upon every $\nu_\text{m}$ clock cycle, the most significant bits of which are used to derive complex NCO phasor components. However, unlike many DDS implementations, SDR often does not use the phase register as an index in a large lookup table of precomputed trigonometric values. Instead, SDR often implements \emph{coordinate rotation digital computer} (CORDIC)~\cite{volder1959cordic} to compute NCO phasors in fixed-point arithmetic. Exploiting the equivalence between angle rotation and phase accumulation, CORDIC is a successive approximation algorithm built from logical operations well suited to a DSP pipeline: comparisons, bit shifts, and binary addition. After inspection of the two most significant accumulator bits fixes the phase quadrant, this SDR implements $K = 20$ CORDIC iterations on 24-bit phase words for an approximate angle resolution of $\tan^{-1} \left[2^{-(K-1)} \right] = 1.9~\mu$rad. CORDIC approximates resampling the real input signal into a complex frame rotating at $f_a$, adding negligible quantization noise (approximately equivalent to $\sigma_x(\tau) = 0.3\,\text{fs } (\tau/1\,\text{s})^{-1/2}$ at 10~MHz). The SDR ultimately truncates the resulting signal to 16 bits of resolution for each of the in-phase ($I$) and quadrature-phase ($Q$) components.

\begin{figure}
\begin{center}
\includegraphics[scale=1]{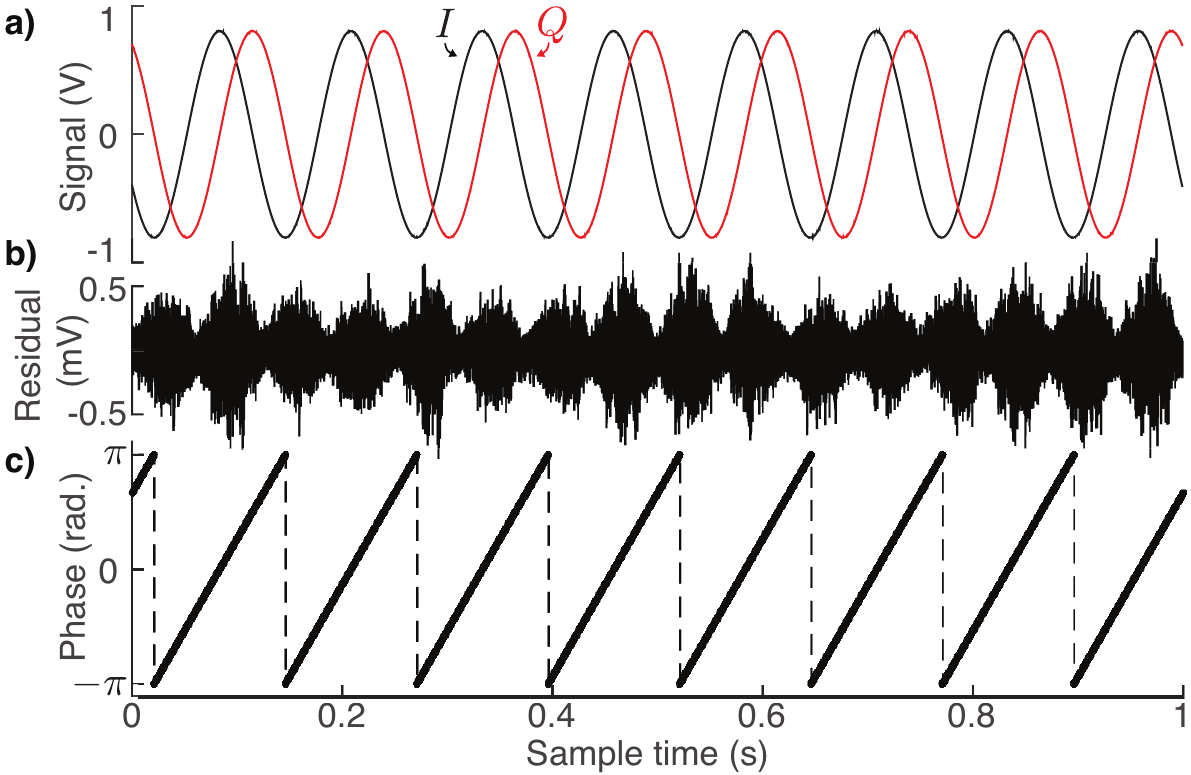}

\end{center}
\caption{SDR measurement of a signal at $f_\text{i} = f_\text{r} =10$~MHz, spectrally-shifted by DDC to $f_\text{b} = 10 \text{ MHz} - (429,496,386 \times 2^{-32} \times 100 \text{ MHz}) \approx 8$~Hz. a) One second ($10^6$ buffered samples) of $z(t_k) = I(t_k) + iQ(t_k)$ data. b) The residual amplitude of $I(t_k)$ after removing a best-fit single-tone. The noise is predominately white, but modulation related to $f_b$ and proportional to $|I(t_k)|$ is clearly observed. c) The instantaneous phase evolves as $2 \pi f_\text{b} t$; here we plot $\arg z(t_k)$ wrapped into $-\pi < \arg z(t_k) \le \pi$.}
\label{fig:waveforms}
\end{figure}

Transmission and manipulation of output samples $z(t_k) = I(t_k) + iQ(t_k)$ at the physical sample rate $\nu_\text{m}$ would require $\approx 3$~Gbps in network and storage resources. Fortunately, DDC shifts the signal of interest close to baseband, allowing aggressive low-pass filtering and rate-decimation (by up to a factor of 512) in hardware. The SDR filters and decimates in three configurable steps. First, a cascaded integrator-comb filter~\cite{hogenauer1981economical} divides the sample rate by an integer $1 \le n_\text{cic} \le 128$. This is followed by two optional half-band decimators~\cite{bellanger1974interpolation}, each accomplishing a rate division of 2 and antialias filtering. Within their passbands these filters have a linear phase/frequency dependence and thus are shape-preserving in the time-domain. Figure~\ref{fig:waveforms} shows data acquired with typical settings, $n_\text{cic}=25$ and both half-band filters enabled, which yields a decimation of $n_\text{dec} = 4 n_\text{cic} = 100$ and $\nu_\text{m}/n_\text{dec} = 10^6$ samples per second, requiring 32 Mbps of network and buffering resources.  ADC quantization noise power, which is nearly uniform in density (`white') over a Nyquist bandwidth of $\pm \nu_\text{m}/2$ is reduced, approximately by $n_\text{dec}^{-1}$ (see Appendix~\ref{sec:decimation_in_practice}). The final DSP section queues and formats $z(t_k)$ along with metadata such as hardware time-stamping and drives their transmission to a general-purpose data acquisition computer. Application programming interfaces are available for several languages, free tools like {\tt gnuradio}~\cite{gnuradio}, and commercial data processing packages.

\section{Demonstration experiments}
We now outline our study of SDR's suitability for oscillator metrology. We first discuss phase measurements over intervals of a few seconds, the analysis of which includes information about fast fluctuations up to $\nu_\text{m} / (2 n_\text{dec})$ in frequency. Then, we consider measurement noise over several hours to days using methods which average over fast fluctuations. We find that over intervals greater than about $10$~ms, ADC aperture jitter is likely a limiting technical noise source. We demonstrate a promising solution available in many SDRs: a second, independent ADC channel is synchronously sampled such that aperture jitter and many other noises subtract in common-mode. We consider application of SDR in a many-clock inter-comparison, and to the problem of optical frequency and phase metrology. Finally, we briefly describe measurement performance of a 6~GHz microwave tone beyond the ADC bandwidth, and the instability of amplitude measurements in two SDR models. 

\subsection{Phase of ADC input vs.\ the sampling timebase}
\begin{figure}
\begin{center}
\includegraphics[scale=0.62]{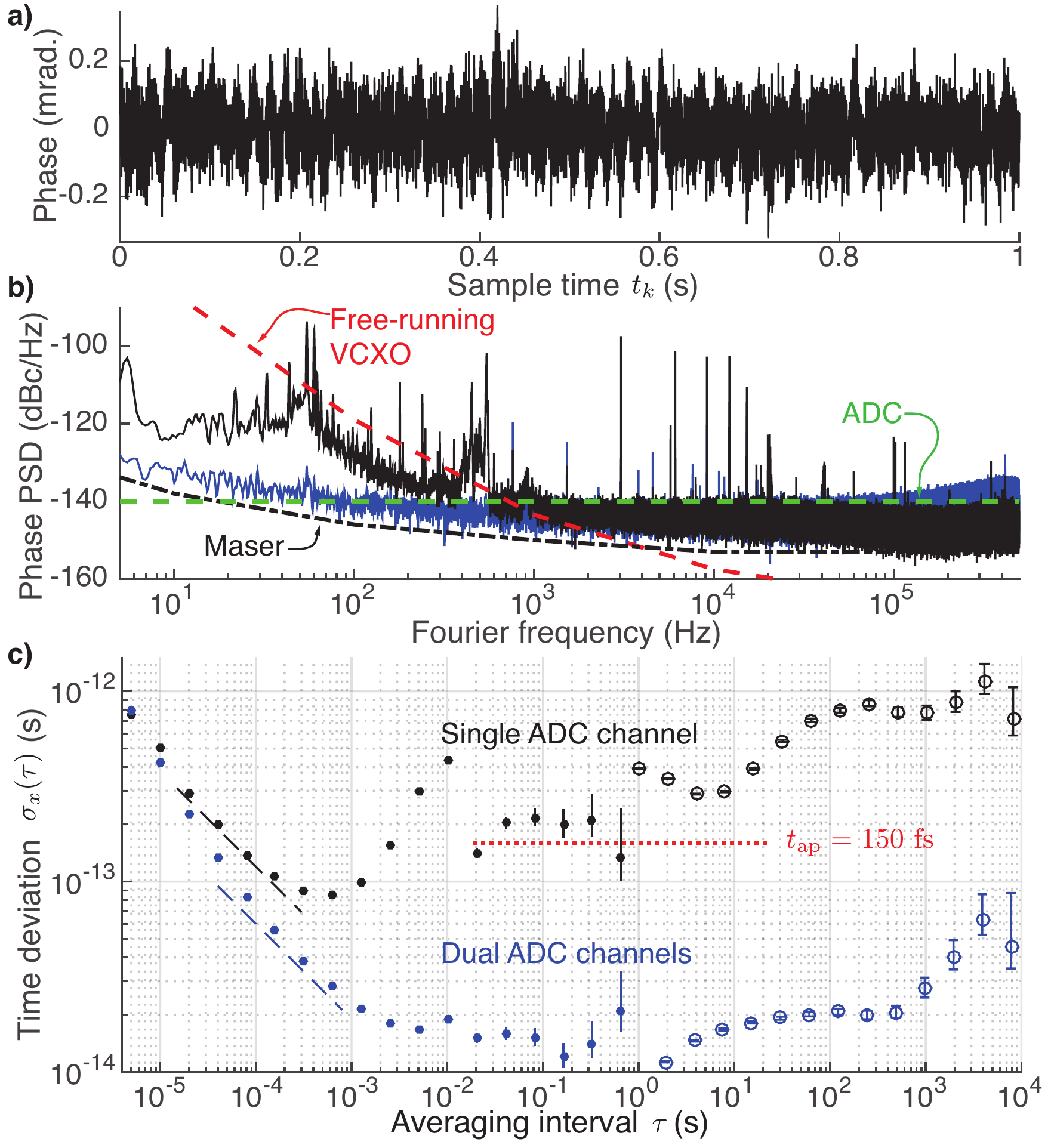}
\end{center}
\caption{a) Unwrapped phase signal $\arg z(t_k)$ when $f_\text{i}$ and $f_\text{r}$ derive from the same 10~MHz oscillator (0.1 mrad corresponds to $\Delta T \approx 1.6$~ps); a deterministic ramp caused by the choice of $f_a$ is removed. Here, $n_\text{dec} = 100$. b) A Fourier transform of $\arg z(t_k)$ (black data) yields a single-sided phase noise power spectral density (PSD). At high Fourier frequencies we observe a white noise of $\approx -140$~dBc/Hz (green dashed line), consistent with the ADC's SNR, signal power, and decimation filtering. At low Fourier frequencies we observe technical noise roughly tracking the rising noise density of the $\nu_\text{m}$ VCXO (red dashed line), which the PLL cannot fully suppress. A hydrogen maser noise specification (black dot-dash) provides context. The relative PSD between two ADC channels (blue data) has much improved flicker noise. c) Time deviation $\sigma_x(\tau)$ in one-channel (black) and two-channel (blue) modes; $n_\text{dec} = 500$. Solid points derive from short streams of $\arg z(t_k)$ samples without averaging. Open circles result from pre-averaging streams in 1~s chunks. White phase noise of $1.2~\text{fs }(\tau/\text{1 s})^{-1/2}$ (black dashed line) is equivalent to $\approx 86$~dB SNR (see Appendix~\ref{sec:spectral_est_phase}). The blue dashed line represents a further 6~dB improvement. A red dashed line marks the ADC's aperture jitter $t_\text{ap}$. For $\tau \gg 10$~s, we expect environmental coupling to dominate both measurement modes. See text for further detail.}
\label{fig:time_psd}
\end{figure}
Consider the arrangement in Figure~\ref{fig:heterodyne_comparison}b where $f_\text{i}$ is approximately known and stationary, and $f_\text{r}$ is treated as a frequency reference ($f_\text{i}$ need not be similar to $f_\text{r}$). We choose the integer $a$ so $f_a$ (see Eq.~\ref{eq:ftw}) is close to $f_\text{i}$. Absent technical noise, $\nu_\text{m} = 10 f_\text{r}$ due to the master timebase's PLL, making $f_a$ exactly computable. The SDR output samples, $z(t_k)$, represent the input signal spectrally shifted to a low frequency $f_\text{b} = f_\text{i} - f_a$; the sample epochs are $t_k = k \times (n_\text{dec}/\nu_\text{m})$. The signal phase, $\arg z(t_k) \equiv \tan^{-1} \left[\text{Im } z(t_k) / \text{Re } z(t_k) \right]$ (see Figure~\ref{fig:waveforms}c) is a time-integral of angular frequency $2\pi f_\text{b}$ and so evolves in time as $2 \pi f_\text{b} t_k + \phi_0$, where $\phi_0$ includes technical offsets such as cable delays. The $\tan^{-1}$ function is evaluated with independent numerator and denominator arguments, removing a phase-quadrant ambiguity. Generally, all SDRs are capable of this mode of measurement, though those without the ability to reference $\nu_\text{m}$ will suffer in accuracy.

To analyze the phase noise floor of this configuration, we split a single 10~MHz oscillator into the $f_\text{i}$ and $f_\text{r}$ inputs. The amplitude at the ADC is kept near half-scale to avoid distortion (typical input power was $\approx 0$~dBm). Ideally, when unwrapped, $\arg z(t_k) = 2 \pi (1 - 10 a/2^{32}) (k n_\text{dec}/10)$ (neglecting a fixed $\phi_0$). In software, we subtract this deterministic trend, the magnitude of which is made small by an appropriate choice of $a$, and interpret residual fluctuations as measurement noise. Figure~\ref{fig:time_psd}a shows a typical residual phase signal, a Fourier transform of which yields the phase noise power spectral density (Figure~\ref{fig:time_psd}b). 

The black curve in Figure~\ref{fig:time_psd}c depicts a complementary statistical measure: the oscillator time deviation~\cite{allan1978time,allan1981modified} $\sigma_x(\tau) = \frac{\tau}{\sqrt{3}} \text{mod } \sigma_y(\tau)$, where $\text{mod } \sigma_y(\tau)$ is the modified Allan deviation~\cite{riley2008handbook}. Briefly, $\sigma_x(\tau)$ characterizes the predictability of phase (in time units, see Eq.~\ref{eq:timephase}) as a function of averaging interval $\tau$. Over roughly $20~\mu\text{s} < \tau < 200~\mu\text{s}$ we observe behavior consistent with white-phase noise, $\sigma_x(\tau) \approx 1.2 \text{ fs } (\tau/1\text{ s})^{-1/2}$. Regrettably, $\sigma_x(\tau)$ stops decreasing with further averaging, and besides an oscillation peak (related to modulation at $f_\text{b}$) appears limited to a flicker-floor roughly consistent with the ADC's $t_\text{ap} = 150$~fs. Reducing input power increases the white-phase instability, but otherwise these performance limits persist over many instrument configurations: rf-coupling method (dc-coupled op-amp vs.\ transformer), choice of heterodyne $f_\text{b}$, decimation factor $n_\text{dec}$, stock vs.\ quiet linear power supply, etc.

\subsection{Phase of one ADC channel vs. another}
\label{sec:two_channel}
To do better we must reduce the influence of phase noise in $\nu_\text{m}$ and the ADC aperture jitter $t_\text{ap}$. Fortunately, many SDRs can process two independent ADC channels which are sampled synchronously (specifically, the two ADC channels exist on the same chip). To examine residual noise in this differential configuration, we split the same oscillator into the three inputs ($f_\text{i}$, $f_\text{j}$, and $f_\text{r}$) as shown in Figure~\ref{fig:heterodyne_comparison}c, though it is not crucial that the $f_\text{r}$ input be identical to either of the others. Since $f_\text{i} = f_\text{j}$, the same NCO frequency $f_a$ is used to DDC both channels, giving the same deterministic trend to both output phase signals. The phase signals should be unwrapped before subtraction because, due to noise and small phase offsets, $2 \pi$-discontinuities can appear at different sample epochs. Example single- and dual-channel signal processing chains are illustrated for comparison in Figure~\ref{fig:heterodyne_comparison}d.

Blue curves in Figures~\ref{fig:time_psd}b and \ref{fig:time_psd}c show the significant improvement in phase noise and time deviation from dual-channel operation. A flicker-floor of $\sigma_x \approx 20$~fs now appears roughly an order-of-magnitude below $t_\text{ap}$ and persists over $1 \text{ ms} < \tau < 500$~s. It also improves by an order-of-magnitude upon the typical noise floor of the DMTD instrument ($\sigma_x \approx 300$~fs). While we lack detailed knowledge of the ADC, we posit that each channel's sample-and-hold circuitry shares a trigger-input threshold-detector. After this element, circuit paths, component/process variation, and environmental non-uniformity on the ADC chip are likely minute. ADC voltage-reference fluctuations and phase noise in the $\nu_\text{m}$ PLL (and its reference, $f_\text{r}$) are also highly common to both sampled channels. Remaining non-common elements include off-chip transmission lines, coupling transformers, and on-chip ADC sampling circuitry. We show later that similar common-mode suppression is present in a different ADC with much larger $t_\text{ap} = 1$~ps.

In this mode of operation, it is less important that $\nu_\text{m}$ be locked to a high-quality oscillator because phase noise in $\nu_\text{m}$ will be highly-common between the two sampled inputs. Noise is not completely suppressed, however. We found slightly better performance, at the level of 20~\% in $\sigma_x$, when $\nu_\text{m}$ was referenced to a hydrogen maser versus the SDR's quartz oscillator. We hypothesize that parasitic coupling of the digital sample clock at $\nu_\text{m}$ is slightly imbalanced between the two ADC inputs. This feature is likely specific to the SDR model and circuit layout.

\subsection{Instability over long averaging intervals}
Maximum decimation in an SDR still results in several megabits per second of data per channel. As a practical matter for long-duration measurements, we reduce this data stream as it is acquired to one recorded $\Delta T$ value per second. This step reduces measurement bandwidth to $\approx 1$~Hz. We tested two simple averaging methods with similar performance: uniform weight (`rectangular window') averaging of $\arg z(t_k)$ over groups of $N = \nu_\text{m}/n_\text{dec}$ samples per second, and the phase estimation routine discussed in Appendix~\ref{sec:spectral_est_phase}. Issues related to windowed averaging here are analogous to those in frequency meters~\cite{rubiola2005measurement}. 

The SDR measurement stability does not degrade much over intervals of several hours, an important requirement for an atomic-clock measurement system~\cite{levine2012invited}. We undertook no special environmental stabilization beyond standard laboratory conditions (ambient temperature control of $\approx 0.5$~K). The SDR operated in its original enclosure with a continuously active cooling fan. For these tests, matched cables were flexible, double-shielded (RD-316), and SMA terminated. Open circles in Figure~\ref{fig:time_psd}c show typical long-term performance of the one- and two-channel SDR techniques. In terms of frequency instability, the two-channel ADC residuals at 10~MHz typically average as flicker-phase noise with $\sigma_y(\tau) = 7 \times 10^{-14}(\tau / \text{1 s})^{-1}$ through $\tau = 10^3$~s.

\subsection{Clock comparison with software radio}
\begin{figure}
\begin{center}
\includegraphics[scale=0.75]{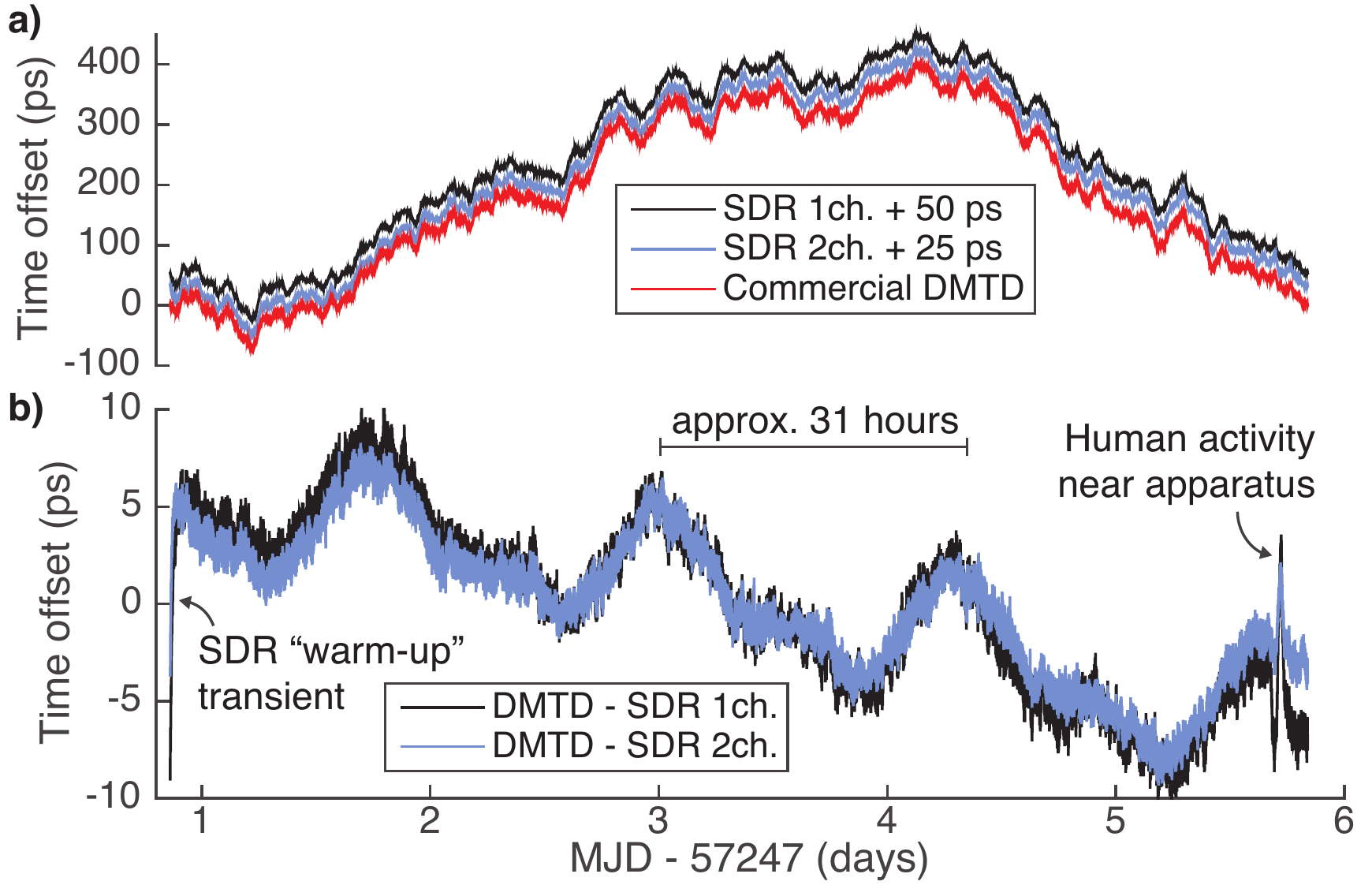}
\end{center}
\caption{a) A comparison of two hydrogen masers over five days (MJD is the modified Julian date), using a commercial instrument based on DMTD (red) and the SDR described here (black/blue for single-/dual-channel mode, respectively). From each time series we subtract a linear phase trend corresponding to the masers' frequency difference of $y = 8.85 \times 10^{-14}$. We introduce 25~ps and 50~ps offsets for visual clarity. b) We show the differences of each SDR measurement with that of the DMTD. Some technical noise features are understood and annotated; it is not yet known whether the DMTD, SDR, or both systems contribute to the $\sim 31$~h periodic modulation.}
\label{fig:maser_compare}
\end{figure}

\begin{figure}
\begin{center}
\includegraphics[scale=0.75]{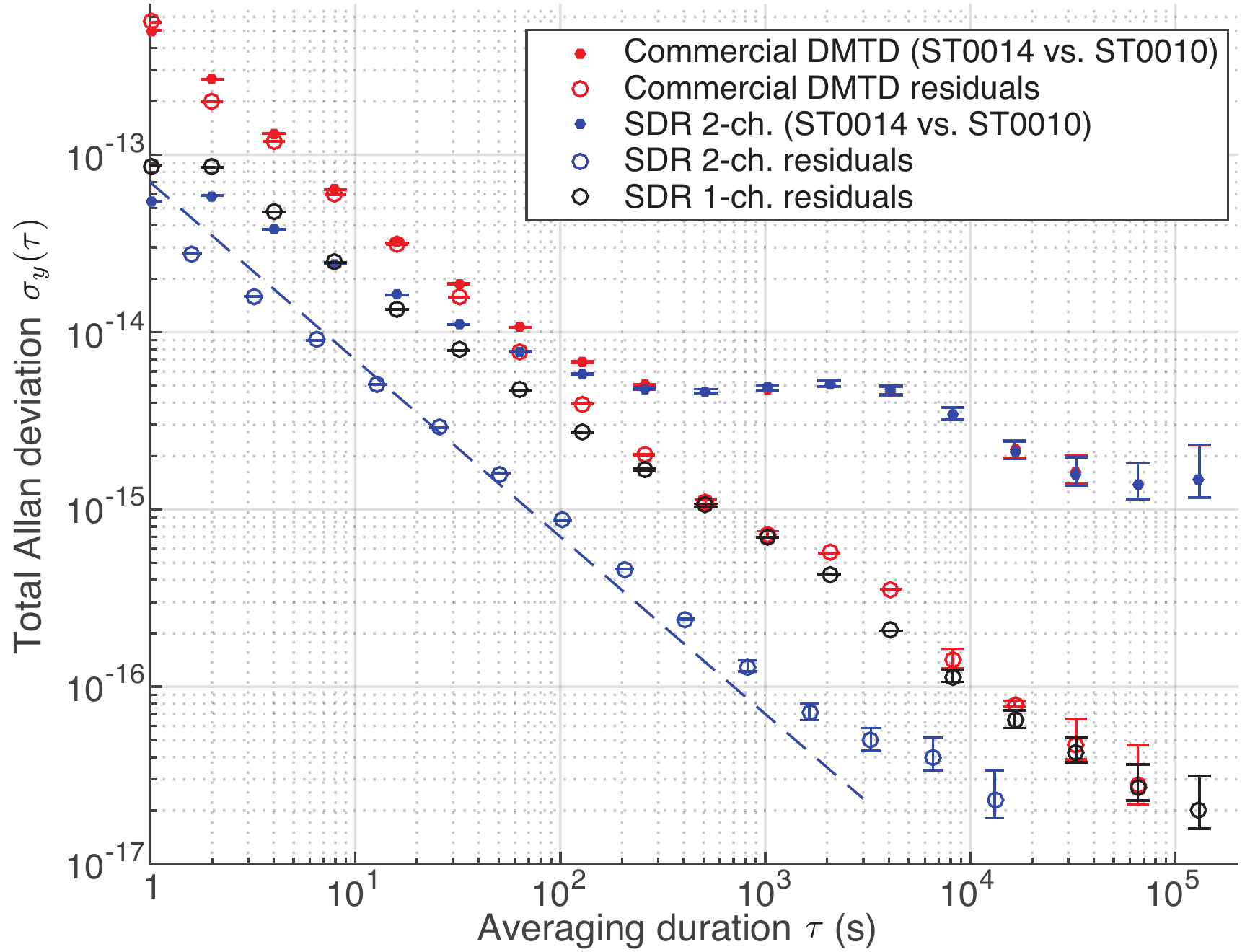}
\end{center}
\caption{Fractional frequency instability~\cite{greenhall1999total} $\sigma_y(\tau)$ of hydrogen masers (NIST masers ST0014 vs.\ ST0010) as measured by a DMTD commercial instrument (red, solid) and the SDR two-channel technique (blue, solid) described here. From $\tau \ge 200$~s both techniques become identically limited by maser frequency fluctuations. Open points show typical residual instabilities of the DMTD instrument (red), the single-channel SDR method (black), and the two-channel SDR method (blue). The blue dashed line is an eye guide placed at $\sigma_y(\tau) = 7 \times 10^{-14} (\tau / \text{1 s})^{-1}.$ Both DMTD and SDR methods yield one datum per second, but the effective measurement bandwidth of the DMTD instrument is known to be $\gg 1$~Hz.}
\label{fig:total_allan}
\end{figure}
The tests described above demonstrate the low instability of the SDR technique; here we discuss time accuracy. Two 5~MHz signals, sourced by hydrogen masers (NIST masers ST0010 and ST0014), are input into the two SDR ADC channels. A non-linear frequency doubler converted one of these to create the $f_\text{r} = 10$~MHz PLL reference. The maser signals were measured simultaneously by a commercial system based on DMTD. Figure~\ref{fig:maser_compare}a shows excellent agreement between the methods. The time-series of the difference between the data sets (Figure~\ref{fig:maser_compare}b) reveals technical noise in one or both measurement systems, some details of which are not yet understood. An initial transient of about 15~ps in magnitude is a repeatable `warm-up' SDR characteristic lasting several minutes. Key component temperatures, measured with platinum resistive thermometers attached with thermally conductive epoxy, increase by 5~K to 10~K in these first several minutes of operation. We also observe a periodic variation (of roughly 31~h) with an amplitude of order 5~ps. Such a variation would contribute $< 10^{-16}$ to fractional frequency instability, which is of marginal significance in the inter-comparison of maser clocks. Figure~\ref{fig:total_allan} shows the frequency instability of the maser comparisons and typical SDR and DMTD residual instabilities.  At averaging intervals of $\tau \approx 10^3$~s, the single-channel SDR technique is comparable with the commercial DMTD instrument; the two-channel SDR technique outperforms both by almost an order-of-magnitude.

\subsection{Multi-channel operation}
\begin{figure}
\begin{center}
\includegraphics[scale=0.75]{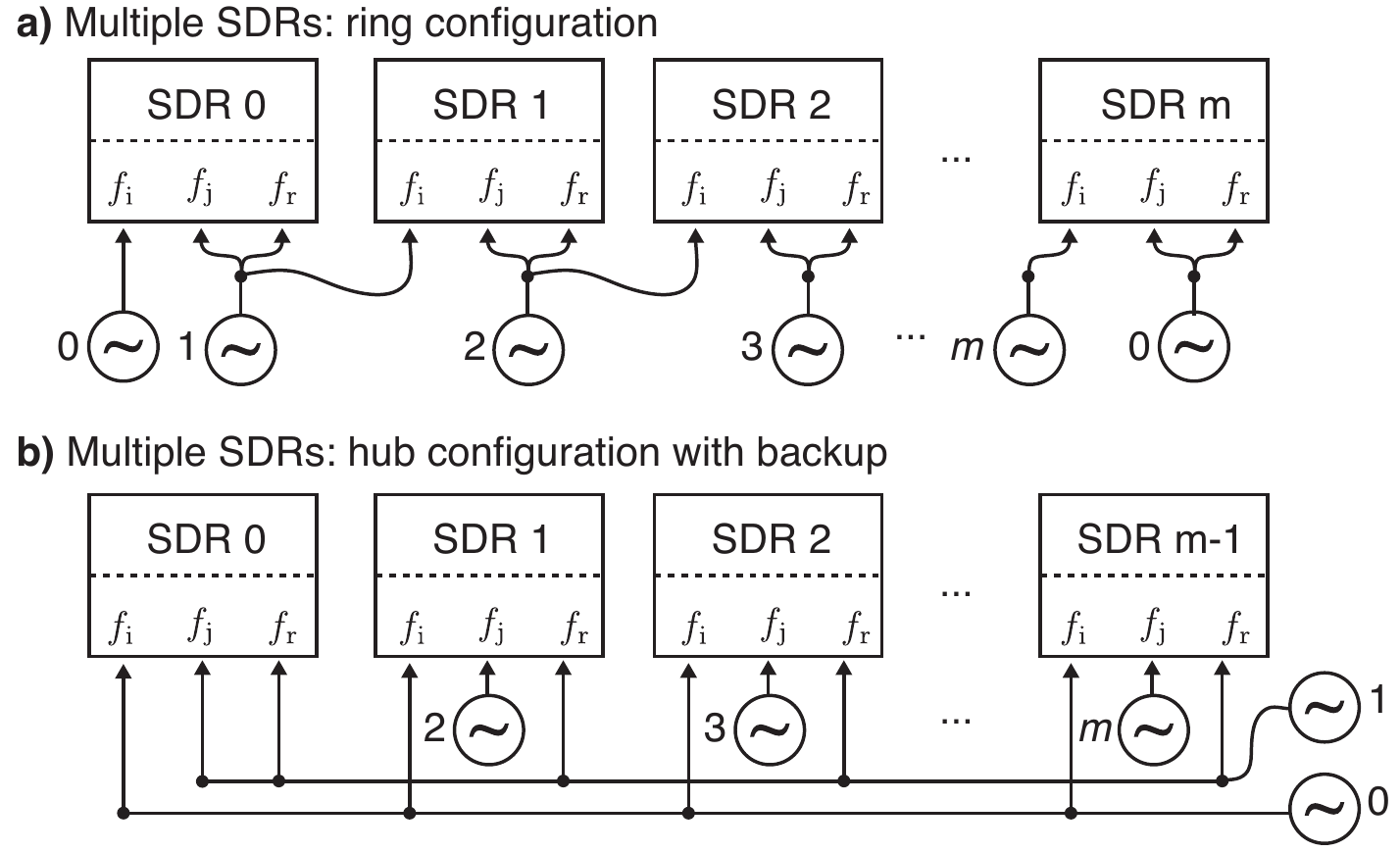}
\end{center}
\caption{Multiple SDRs can scale for coherent many-oscillator comparisons in flexible arrangements, the choice of which will depend on which failure modes are judged most likely. a) For example, in a `ring' configuration, each SDR node produces the two-channel differential signal $f_\text{j} - f_\text{i}$, a one-channel signal $f_\text{i} - f_\text{r}$ and unique one-channel residual $f_\text{j} - f_\text{r}$. Phase data collection for all oscillators is uninterrupted with any single node failure. b) In a `hub' configuration, the oscillator indexed `0' is distributed to an ADC channel in each node as part of a two-channel differential measurement. To protect the network against failure of oscillator `0', oscillator `1' provides a shared PLL reference to all nodes, enabling one-channel measurements of all oscillators as a `degraded backup'. Here, junctions imply distinct distribution amplifier channels; differential amplifier and cable delays must be accounted for when comparing oscillator phase differences.}
\label{fig:multi_sdr_configs}
\end{figure}
Some commercial DMTD instruments accept 16 or more input oscillators, where one channel is permanently assigned a special role as reference for the TIC timebase $\nu_\text{m}$. The two-channel SDR scheme presented here is scalable to an unlimited number of channels, and it is possible but not necessary that one oscillator be assigned a special role. Figure~\ref{fig:multi_sdr_configs}a sketches a scheme whereby multiple SDR instruments are arranged in a `ring,' immune to any single oscillator or SDR fault. A `hub' model (Figure~\ref{fig:multi_sdr_configs}b), where one oscillator is distributed to all measurement nodes is also possible. Simultaneous implementation of the one-channel SDR technique using a distinct $f_\text{r}$ oscillator provides a `backup hub' mode of operation with degraded performance. In a scaled deployment, it may be desirable to increase the decimation performed in hardware, perform the phase computation and averaging itself in the FPGA, and/or distribute the software data processing among multiple connected computers. We estimate that, per measurement channel, the material cost of a SDR solution is a factor of two or more below competitive multi-channel DMTD instruments.


\subsection{Optical oscillator measurement}
\label{sec:optical}
\begin{figure}
\begin{center}
\includegraphics[scale=0.65]{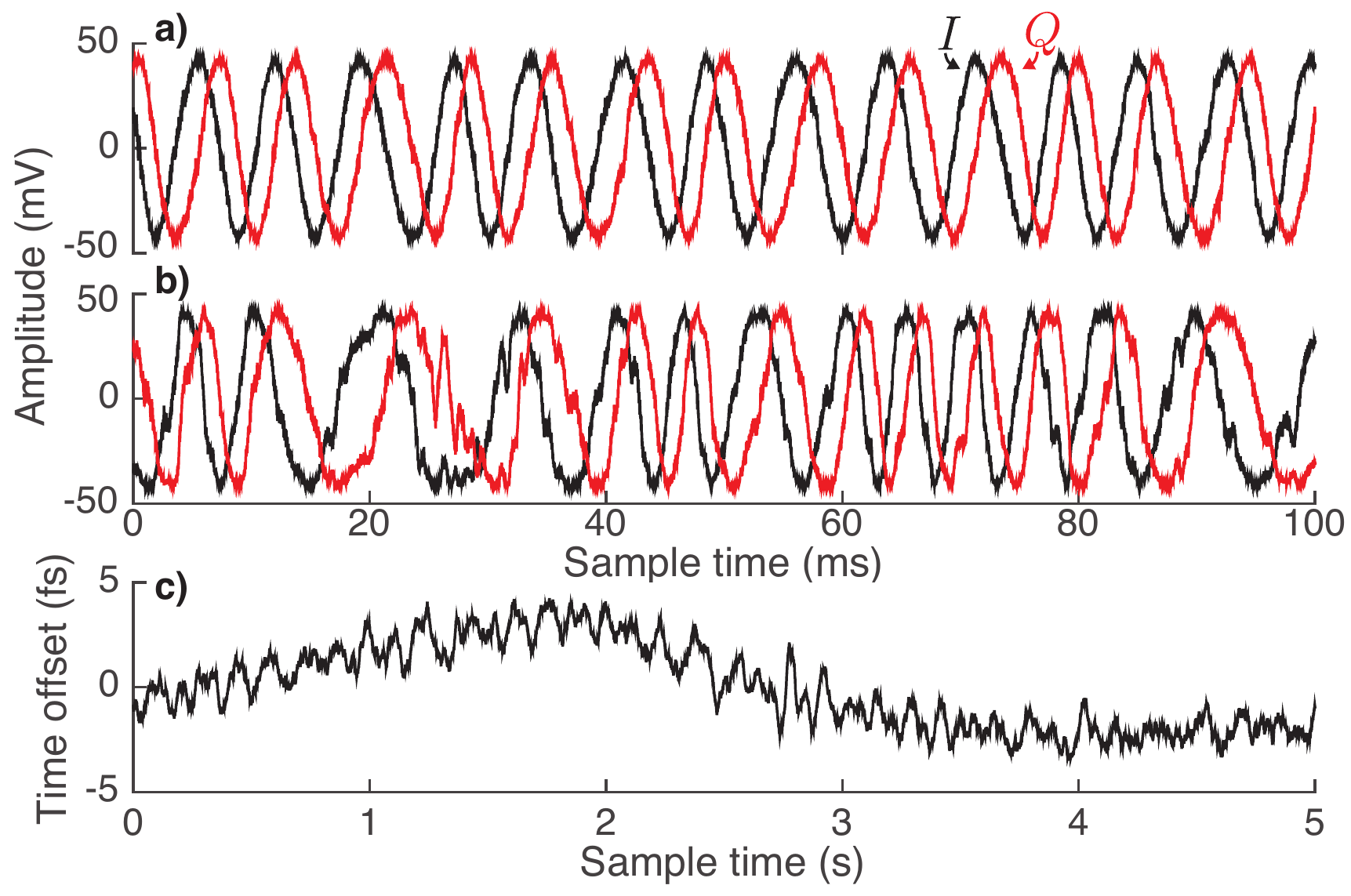}
\caption{Tracking optical phase with SDR (see text for details). a) The SDR down-converts a heterodyne between a femtosecond laser frequency comb (FLFC) stabilized to reference laser $f_\alpha$, and laser $f_\beta$ to an audio tone of $\approx 140$~Hz. We plot the complex components of the SDR output $z(t_k)$. b) Laser $\beta$ is transmitted to the FLFC heterodyne via an uncompensated fiber optic link. By shaking the fiber, we observe and can coherently track resulting phase fluctuations. c) Dividing $\arg z(t_k)$ by $2 \pi f_\beta$, we cast phase fluctuations as time instability of the optical oscillator $\beta$. A constant phase and frequency offset are suppressed in the plot.}
\label{fig:optical_time}
\end{center}
\end{figure}
Optical atomic frequency references now exceed the performance of official primary standards based on microwave frequencies by factors of 1000 in stability~\cite{hinkley2013atomic} and potentially 100 in accuracy~\cite{heavner2014first,ushijima2015cryogenic,bloom2014optical}. Generally, optical frequency references operate by disciplining a pre-stabilized laser oscillator to an atomic resonance in neutral atoms or single trapped ions~\cite{poli2013optical}. Direct phase and frequency comparisons between two lasers at $f_\alpha$ and $f_\beta$ are only possible if they are sufficiently close to create a heterodyne beatnote on a photodiode or other transducer. Otherwise, a now standard technique employs a broadband femtosecond laser frequency comb (FLFC, or comb) as a common heterodyne oscillator spanning hundreds of terahertz~\cite{udem2002optical,newbury2011searching}. A FLFC spectrum consists of many optical modes, whose absolute frequencies can be expressed as $f_n = n f_\text{rep} + f_\text{ceo}$ for many thousands of consecutive integers $n$. The comb's pulse repetition rate, $f_\text{rep}$, scales inversely with the laser resonator length, and $|f_\text{ceo}| < f_\text{rep}$ depends on the details of the intra-cavity dispersion. For our purposes, it is sufficient to note that both degrees of freedom correspond to radio frequencies controllable by phase-lock techniques.

We measured and tracked phase fluctuations between two laser oscillators using a FLFC and the SDR. A Ti:sapphire FLFC~\cite{fortier2006octave} with $f_\text{rep} \approx 1$~GHz was stabilized by locking a comb mode $640$~MHz offset from an ultra-stable optical frequency $f_\alpha \approx 259$~THz. We used self-referencing interferometry~\cite{cundiff2003colloquium} to stabilize $f_\text{ceo}$. A second ultra-stable laser~\cite{fortier2015digital}, $f_\beta \approx 282$~THz, interfered with another comb mode to make a heterodyne tone $f_\text{o}$ near 160~MHz on an amplified photodiode. Independent characterizations have determined frequency instability floors of $\le 2 \times 10^{-16}$ for laser $\alpha$ and $1 \times 10^{-15}$ for laser $\beta$. Due to the comb's phase locks, fluctuations of $f_\text{o}$ are directly related to the fluctuations between the $\alpha$ and $\beta$ laser oscillators; the required comb mode integers for absolute determinations can be obtained by low-resolution wavemeter measurements of $f_\alpha$ and $f_\beta$. 

Traditionally, only gated frequency measurements are made of $f_\text{o}$, discarding information about phase fluctuations. A DMTD scheme to track phase is impractical: generally, $f_\text{o}$ can appear at any frequency up to $f_\text{rep}/2$ depending on FLFC preparation, and $f_\text{o}$ fluctuations and drift are typically too large. In contrast, the SDR has a high input bandwidth, a tunable NCO for down-converting arbitrary $f_\text{o}$, and tracks phase information over very short intervals $\nu_\text{m}/n_\text{dec} \le 5~\mu$s with no dead time.

Since the ADC sample clock $\nu_\text{m} = 100$~MHz, $f_\text{o} \approx 160$~MHz appears in the third $\pm \nu_\text{m}/2$ Nyquist zone, aliased to $-40.005~860$~MHz. We set the NCO $f_a = -40.006~000$~MHz in order to obtain an audio beat note $|f_\text{b}| \approx 140$~Hz. Figure~\ref{fig:optical_time}a shows the output sample data under normal conditions; Figure~\ref{fig:optical_time}b shows directly observable phase noise created by vigorously shaking the uncompensated~\cite{ma1994delivering} fiber optics coupling laser $\beta$ to the comb. It is important to appreciate that a radian of optical phase remains unscaled by mixing with the comb to make $f_\text{o}$, nor is it scaled by the DDC process $f_\text{o} \to f_\text{b}$ in the SDR. So, treating laser $\alpha$ as a reference, we can derive the time fluctuations of laser $\beta$ by unwrapping and dividing the $f_\text{b}$ phase $\arg z(t_k)$ by a factor $2 \pi \times 282$~THz, following Eq.~\ref{eq:timephase}. Figure~\ref{fig:optical_time}c shows the result: well-resolved femtosecond-level temporal instability between two would-be optical clocks, lasers $\alpha$ and $\beta$. In measurement of optical heterodyne tones, the SDR noise floor is negligible.

A multi-channel SDR arrangement monitoring several FLFC heterodyne beat notes could form the measurement basis for an \emph{optical time scale}, meaning an ensemble of optical oscillators statistically weighted to produce a robust and reliable `average clock'~\cite{levine2012invited,tavella1991comparative}. Related technology is approaching a high level of readiness, including robust fiber-FLFC designs~\cite{sinclair2014operation}, stabilized `flywheel' lasers~\cite{hafner20158, kessler2012sub} with frequency instabilities $\sigma_y \le 1 \times 10^{-16}$, and optical atomic standards characterized at the $10^{-18}$ uncertainty level~\cite{bloom2014optical}.

\subsection{Microwave frequencies}
\label{sec:microwave}
\begin{figure*}
\begin{center}
\includegraphics[scale=0.75]{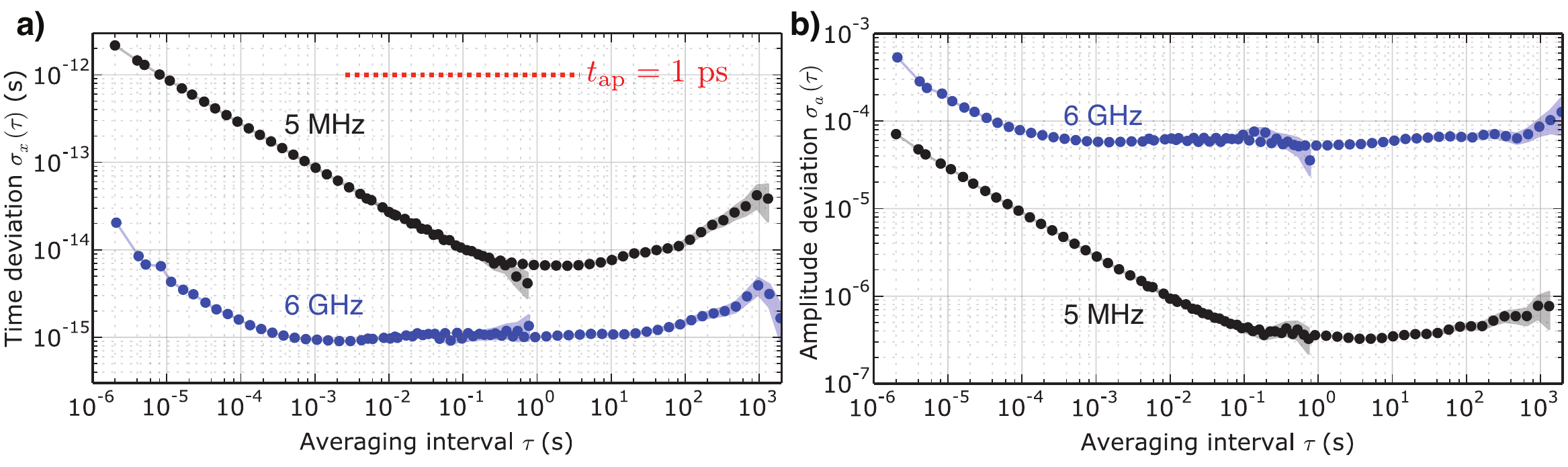}
\caption{a) Time deviation of differential phase measurements of 5 MHz and 6 GHz signals. For 5~MHz, the SDR featured a 12-bit ADC with 1~ps aperture jitter. The instability floor is more than two orders of magnitude lower than $t_\text{ap}$, indicating excellent common-mode suppression of technical noise. For 6~GHz, the SDR featured a LO synthesizer and analog mixer front-end to translate the signal into the ADC bandwidth. Though phase-noise performance is made worse by these elements, the high signal frequency leads to a time stability floor of 1~fs, roughly an order-of-magnitude better than the results at 10~MHz (Figure~\ref{fig:time_psd}c). b) We also investigated amplitude measurement instability (normalized to input amplitude) of two-channel signals in these SDR models. In both plots, data at longer $\tau$ are obtained by additional software decimation by a factor of 2500 prior to storage. These data were acquired in an unstabilized office environment and with the sample clock $\nu_\text{m}$ un-referenced. Shaded bands indicate standard statistical uncertainties.}
\label{fig:microwave_and_amplitude}
\end{center}
\end{figure*}

Microwave frequencies far beyond the ADC input bandwidth are measurable by SDR models that incorporate an analog mixer and microwave local-oscillator (LO) synthesizer referenced to the same source as $\nu_\text{m}$. In a separate investigation, we tested a SDR (Ettus USRP B210) featuring such a front-end (Analog Devices AD9361) capable of down-converting two $\le 6$~GHz signals before sampling them at 12-bit resolution. To characterize its phase-noise performance, we input a 6~GHz ($-22$~dBm) signal and set the SDR's programmable amplifiers to 49~dB to use the full ADC range. $\nu_\text{m}$ was set to 30.72~MHz, and $n_\text{dec}$ to 32. Due to the mixer front-end, we observed significantly higher phase noise than the results in section~\ref{sec:two_channel}: a white noise floor at $-123\,\text{dBc/Hz}$ and flicker noise of $-90\,\text{dBc/Hz}~(f/1\,\text{Hz})^{-1}$. However, given the much higher carrier frequency, the equivalent time deviation limits were $\sigma_x(\tau) = 20\,\text{as}~(\tau/1\,\text{s})^{-1/2}$ over short intervals and a flicker floor of $1\,\text{fs}$, as shown in Figure~\ref{fig:microwave_and_amplitude}a.

\subsection{Amplitude metrology}
\label{sec:amplitude}
Though we have so far ignored it, the amplitude of a complex sampled SDR signal is also available as $\sqrt{I^2(t_k) + Q^2(t_k)}$. In a separate investigation, we studied the relative amplitude instability limit of signals input into two ADC channels. We tested a SDR (Ettus USRP B100) with a 12-bit ADC (Analog Devices AD9862), $\nu_\text{m} = 64$~MHz, $n_\text{dec} = 64$, and $f_\text{i} = f_\text{j} = 5$~MHz. As shown in Figure~\ref{fig:microwave_and_amplitude}b, we observed a relative amplitude instability floor of $3 \times 10^{-7}$ over the averaging interval $0.1~\text{s} \le \tau \le 100~\text{s}$. The 6~GHz configuration, described in section~\ref{sec:microwave}, achieved an amplitude instability floor of $5 \times 10^{-5}$. 

\section{Conclusions}
Generally, SDR receivers are little more than high-speed signal samplers followed by a series of digital filters designed to reduce data rate and noise bandwidth. However, these few ingredients are sufficient for several recipes in high-precision time and frequency metrology. Phase/time-offset measurements using unmodified SDR hardware can exceed the stability performance of a commercially-available instrument based on the classic DMTD design while offering increased flexibility. SDR measurement of phase using two input channels differentially reduces the influence of technical timing noise and has demonstrated a maser clock frequency resolution $\sigma_y \le 10^{-16}$ within $10^{3}$~s of averaging. Over several days of continuous hydrogen maser measurement, the SDR technique appears highly accurate, with relatively low environmental noise coupling in a typical laboratory environment. SDR hardware is scalable to coherently measure any number of oscillators at almost any radio or microwave frequency. We have shown the SDR can resolve relative oscillator amplitude fluctuations below the part-per-million level. Finally, we have demonstrated that SDR can be usefully employed in the comparison of ultra-stable optical clocks and oscillators by measuring heterodyne products of clocks with a femtosecond laser frequency comb.

Useful extensions of this work could include a long-term frequency comparison of atomic-clocks' output signals at multiple frequencies (e.g., 5 MHz and 100 MHz), and integration of a many-channel fast ADC into an SDR architecture for better multi-channel scalability. Alternatively, the transmission functions of the SDR could be employed in active phase-noise compensation in optical or FLFC interferometry applications.

NIST's Time and Frequency Division funded this investigation. The work is a contribution of NIST and not subject to U.S.\ copyright. The authors thank Judah Levine for helpful discussions, Joshua Savory (maser comparisons), Franklyn Quinlan (FLFC measurements), and Roger Brown for careful reading of the manuscript. 

\appendix
\section{Spectral estimation of frequency and phase}
\label{sec:spectral_est_phase}
In the single-channel setup of Figure~\ref{fig:heterodyne_comparison}b, a $f_\text{i}$ known only to within a Nyquist bandwidth $\nu_\text{m}/2$ can be quickly acquired by seeking high spectral power while scanning the NCO $f_a$ over its full range. Without loss of generality, we suppose $f_\text{i} < \nu_\text{m}/2$ and choose the calculable NCO frequency $f_{a}$ such that $|f_\text{i} - f_a| \ll \nu_\text{m}/n_\text{dec}$; in other words, the DDC frequency must be within the decimated Nyquist zone. The sign of the sampled `beatnote' $f_\text{b} = f_\text{i} - f_a$ is fixed by the sense of temporal phase rotation in $z(t_k)$, or equivalently, the phase relationship of its real and imaginary components. The problem of high resolution determination of $f_\text{i}$ reduces to spectral estimation on groups of $N$ samples of $z(t_k)$ to estimate $f_\text{b}$. Though no closed-form solution exists generally for spectral estimation~\cite{kaymodern}, our circumstances are unusually favorable: $z(t_k)$ consists of a single, low-frequency tone $f_\text{b}$, with high SNR and little harmonic distortion. Though computationally intensive, an optimal un-biased frequency estimator given these assumptions is the argument $\hat{f}_\text{b}$ maximizing the basic periodogram function~\cite{stoica2005spectral} $|P(f)|^2$, where
\begin{equation}
\label{eq:periodogram}
P(f) = \frac{1}{N} \sum_{k=0}^{N-1} z(t_k) e^{-i 2 \pi f t_k}.
\end{equation}
For signals like ours, $|P(f)|$ is well-approximated by a quadratic polynomial near its maximum. We therefore implemented Brent's method of one-dimension parabolic interpolation~\cite{press1996numerical} to efficiently search for $\hat{f}_\text{b}$. A lower resolution FFT-based spectral estimator seeds this non-linear search with an initial guess. Unlike such FFT-based methods, no windowing function or zero-padding must be applied to the sampled data prior to $|P(f)|^2$ maximization, and there is no need to make $N$ a power-of-2. The search also yields an optimal estimator for the single-tone amplitude, $\hat{A}_\text{b} = |P(\hat{f}_\text{b})|$. In the limit of high SNR and spectrally-uniform uncorrelated (`white') noise, periodogram maximizing spectral estimates converge with maximum likelihood and non-linear least-squares fit results. 

Figure~\ref{fig:total_allan} (black open circles) shows measured frequency instability of a $f_\text{i} = f_\text{r} = 10$~MHz signal, which surpasses that of a commercial frequency meter of comparable cost. Importantly, note that the SDR measurement instability decreases as $\tau^{-1}$, compared to many frequency meters' instability $\propto \tau^{-1/2}$. This difference is attributable to non-zero dead-time and frequency quantization in commercial meter readings. The interval $N (\nu_\text{m}/n_\text{dec})^{-1}$ is analogous to a `gate interval' of a traditional frequency meter. With SDR, this parameter may be chosen during or after data acquisition since $z(t_k)$ data can be stored. Barring interruption in data transmission, this method of frequency analysis has zero `dead-time' intervals during which the input oscillations are unmeasured.

We continue the spectral estimation method to determine phase offset measurements from sets of $N$ complex waveform samples $z(t_k)$. If unknown, we first find the $\hat{f}_\text{b}$ maximizing the periodogram function $|P(f)|^2$ from Eq.~\ref{eq:periodogram}. Then, the optimal estimate of the signal's phase is
\begin{equation}
\hat{\phi}_\text{b} = \tan^{-1} \left[ \frac{\text{Im }P(\hat{f}_\text{b})}{\text{Re }P(\hat{f}_\text{b})} \right].
\end{equation}
Successive estimates of phase on continuously sampled data will evolve as
\begin{align}
\hat{\phi}_\text{b}(t_k) &= \phi_0 + 2 \pi f_\text{b} t_k \\
 			&= \phi_0 + 2 \pi (f_\text{i} - f_\text{r}) t_k + 2 \pi f_\text{r} \left( 1 - \frac{\nu_\text{m}}{f_\text{r}} \frac{a}{2^{32}} \right) t_k,
\end{align}
where $\phi_0$ is the initial phase offset and, for the SDR described in section~\ref{sec:technical_details}, $\nu_\text{m}/f_\text{r} = 10$. The final term, the result of our choosing a heterodyne offset frequency, is exactly computable in terms of $f_\text{r}$ and is removable in post-processing. Subtracting it using complex phase rotation neatly avoids $2\pi$ discontinuities, leaving us only with a phase growing linearly with the frequency difference of interest $f_\text{i} - f_\text{r}$. Phase discontinuities must still be expected and handled over time intervals $\tau \ge 2 \pi/(f_\text{i}-f_\text{r})$. The variance of a single $\hat{\phi}_\text{b}$ estimate using $N \gg 1$ samples is bounded by~\cite{kaymodern}
\begin{equation}
\text{var} \left( \hat{\phi}_\text{b} \right) \ge \frac{1}{\text{SNR}} \frac{2}{N}.
\end{equation}
As $\text{SNR} \propto 1/N$ itself (due to process gain), the bound for variance in the phase estimator is independent of the sample density $N$ under optimal noise conditions, remaining inversely proportional to the SNR and total observation duration. Combining this result with Eq.~\ref{eq:timephase}, the resulting theoretical bound on  time deviation~\cite{sullivan1990characterization} is
\begin{equation}
\label{eq:timephase_noiselimit}
\sigma_x(\tau) = \frac{1}{2 \pi f_\text{i}} \sqrt{\text{var}(\hat{\phi_\text{b}})} \ge 1.2 \times 10^{-15} \text{ s} \, \, (\tau/\text{1 s})^{-1/2},
\end{equation}
where $f_\text{i} = 10$~MHz, $N = 10^6$ samples per second, and the effective SNR $\approx 86$~dB (see Appendix~\ref{sec:decimation_in_practice}). Observations in Figure~\ref{fig:time_psd} (black solid points) are consistent with this noise limit over short averaging intervals.

\section{Decimation fidelity in practice} \label{sec:decimation_in_practice}
\begin{table}
\begin{center}
\caption{Decimating low-pass filters in the SDR ideally improve SNR proportionally to the decimation factor $n_\text{dec}$. We a slightly worse empirical scaling  $\propto n_\text{dec}^{0.8}$. Here we compare the measured SNR for a constant, half-scale, $f_i = f_r = 10$~MHz maser-referenced tone under different decimation settings. $f_b \approx 8$~Hz; other choices yielded similar results.}
\begin{tabular}{rccc} \hline \hline
$n_\text{dec}$	& Expected SNR (dB) 	& Observed SNR (dB)	& Excess noise (dB) \\ \hline
20			& 81.5		& 74.5		& 7.0 \\
40			& 84.5		& 76.7		& 7.8 \\
100			& 88.5		& 79.8		& 8.7 \\
200			& 91.5		& 82.1		& 9.4 \\
500			& 95.5		& 85.2		& 10.3 \\ \hline
\end{tabular}
\label{tab:excess_noise}
\end{center}
\end{table}
Ideally, in the presence of uniform Gaussian noise, the SNR of ADC samples should be improved by a factor of the decimation ratio $n_\text{dec}$ since the CIC and half-band decimating filters approximate an ideal low-pass filter. Alternatively, with SNR is expressed in dB,
\begin{equation}
\text{SNR}_\text{(ideally observed)} = \text{SNR}_\text{ADC} + 10 \log n_\text{dec}.
\end{equation}
However, finite precision in the numerical filters, and the presence of non-Gaussian noise, such as spurs and input noise near the sample clock $\nu_m$, result in slightly worse performance. We observe an approximate $n_\text{dec}^{0.8}$ improvement with $20 \le n_\text{dec} \le 500$ as shown in Table~\ref{tab:excess_noise}.

\bibliography{sdr_tf}

\begin{thebibliography}{10}
\bibitem{nist_disclaimer}
Certain commercial equipment, instruments, or materials are identified in this
  paper for informational purposes only. Such identification does not imply
  recommendation or endorsement by the National Institute of Standards and
  Technology, nor does it imply that the materials or equipment identified are
  necessarily the best available for the purpose.
\end{thebibliography}

\end{document}